\documentclass[12pt,english,floatfix,superscriptaddress,aps,preprint,amsmath,amssymb]{revtex4}
\usepackage[latin9]{inputenc}
\usepackage{amsmath}
\usepackage{amssymb} % For \hslash
\usepackage{amsbsy}
\usepackage{amsfonts}
\usepackage{amsopn}
\usepackage{amstext}
\usepackage{graphicx}
\usepackage{esint}
\usepackage{hyperref}

\usepackage{amssymb}
\usepackage{amsfonts}
\usepackage{amsmath}
\usepackage{graphicx}
\usepackage[english]{babel}
\usepackage{color}
\usepackage[dvips]{epsfig}
\usepackage[dvips]{graphicx}
\usepackage{float}
\usepackage{units}
\usepackage{textcomp}
\usepackage{babel}

\def \be {\begin{equation}}
\def \ee {\end{equation}}
\def \bea{\begin{eqnarray}}
\def \eea{\end{eqnarray}}
\def \ba {\begin{align}}
\def \ea {\end{align}}

\def \a {\alpha}
\def \b {\beta}

\def \G {\Gamma}
\def \d {\delta}

\def \m {\mu}
\def \n {\nu}

\def \l {\lambda}

\def \s {\sigma}

\def \r {\rho}
\def \o {\omega}
\def \O {\Omega}

\def \p {\partial}
\def \f {\frac}

\def \nn {\nonumber}

\def \la {\leftarrow}

\def \la {\label}

\begin{document}

\title{Induced geometry from disformal transformation}

\author{Fang-Fang Yuan}
\email{ffyuan@nankai.edu.cn}
\affiliation{School of Physics, Nankai University, Tianjin 300071, China}

\author{Peng Huang}
\email{huangp46@mail.sysu.edu.cn}
\affiliation{School of Astronomy and Space Science, Sun Yat-Sen University, Guangzhou
510275, China}

\begin{abstract}

In this note, we use the disformal transformation to induce a geometry from the manifold which is originally Riemannian. The new geometry obtained here can be considered as a generalization of Weyl integrable geometry. Based on these results, we further propose a geometry  which is naturally a generalization of Weyl geometry.

\end{abstract}
\maketitle

\section{Introduction}

Disformal transformation was first introduced by Bekenstein as a generalization of usual conformal transformation in theories of gravity \cite{Bekenstein:1992pj}.
Unlike the conformal transformation where only a positive definite functional of scalar field is involved,
the covariant derivatives of scalar field are also needed to define the disformal transformation of metric.
This kind of transformation has been applied to various areas including varying speed of light models \cite{Bassett:2000wj},
inflation \cite{Kaloper:2003yf},
relativistic extensions of modified Newtonian dynamics paradigm \cite{Bekenstein:2004ne},
dark energy models \cite{Zumalacarregui:2010wj} and dark matter models \cite{{Bettoni:2011fs},{Chamseddine:2013kea},{Deruelle:2014eha}}.
Furthermore, as generalized scalar-tensor theories with second order field equations,
Horndeski theories have been extensively studied in recent years \cite{{Horndeski:1974wa},{Deffayet:2013lga},{Deffayet:2011gz}}(for even more general, but healthy, scalar-tensor theories and their further important developments, see \cite{{Gleyzes:2014dya},{Gao:2014soa},{Gao:2014fra}}). Interestingly, they are shown to be invariant under a special class of disformal transformations  \cite{Bettoni:2013diz}. By introducing some constraints, this result can be extended \cite{Zumalacarregui:2013pma}.
For a sample of recent developments concerning disformal transformation, see \cite{{Koivisto:2012za},{Goulart:2013laa},{Brax:2013nsa},
{Akrami:2014lja},{Brax:2014vva},{Creminelli:2014wna},{Minamitsuji:2014waa},{Sakstein:2014isa},{Sakstein:2014aca},{Tsujikawa:2014uza},{vandeBruck:2015ida}}.

On the other hand, it is known for a long time that Weyl integrable geometry \cite{Weyl:1918ib} has intimate relations with conformal transformation
\cite{{Novello:2008ra},{Scholz:2011za}}.
Accordingly, in the context of scalar-tensor theories of gravity including Brans-Dicke theory,
the frames issue and scale invariance issue have always been the focus of debate.
A new viewpoint introduced
by Quiros et al. \cite{{Quiros:2011wb},{Quiros:2014wda}} is that
the resolution of these physical questions
depends on how we assign the affine structure to the underlying spacetimes.
A strict equivalence between Jordan's frame (JF) and Einstein's frame (EF) requires us precisely incorporate the Riemannian structure of the starting spacetime (in which the scalar-tensor theory lives) through conformal transformation into the new geometry, and this operation will inevitably cause the initial Riemannian spacetime to change into Weyl integrable geometry after conformal transformation.
This point of view takes advantage of a well-known fact that Weyl integrable geometry and Riemannian geometry can be transformed into each other by appropriate Weyl rescalings, thus they in fact describe the same spacetime but with different gauge, see \cite{Romero:2012hs} for a well demonstration on this issue.

In history, Weyl integrable geometry was proposed to get rid of the second clock effect, and its tight connection to Riemannian geometry comes from careful investigation of itself. However, as will be show evidently in Section \ref{sec3}, Weyl integrable geometry can be induced from Riemannian geometry by conformal transformation entirely without any \emph{a priori} knowledge of the so-called Weyl integrable geometry. With this observation in mind, a natural question to ask is: what new geometry will be induced if we implement disformal instead of conformal transformation to the metric in the original Riemannian geometry? A clear answer to this question may bring benefits in twofolds. The most direct benefit is, mathematically, it will give us new geometry whose importance needs further study.   Physically, it may help to understand the equivalence between disformal frames and EFs in Horndeski theories  \cite{{Horndeski:1974wa},{Deffayet:2013lga}}, just as what has happened in usual scalar-tensor theories, and this is important for our understanding of the most general scalar-tensor theories. Thus, in this work, we would like to investigate this problem and figure out what geometry will be induced from disformal transformation.

The outline of this paper is as follows.
In Section \ref{sec2}, we review the conceptional basis of Weyl and Weyl integrable geometry in some detail.
In Section \ref{sec3}, we firstly show in detail how to induce Weyl integrable geometry from Riemannian geometry by conformal transformation, from which we extract a general and explicit strategy for our further investigation. After the foundation is laid, we then focus on the special disformal transformation
and obtain a new induced connection.
The corresponding gauge transformation is also derived.
We conclude our discussions in the last section.

\section{A brief review of Weyl integrable geometry}   \la{sec2}

Weyl geometry was proposed by Weyl in 1918 as an attempt to unify gravity with electromagnetism \cite{Weyl:1918ib}.
While the metricity condition of Riemannian geometry reads as
$\nabla_{\m}g_{\a\b}=0$, the Weyl nonmetricity condition reads
\be
\label{2.1}
^{(W)}\nabla_{\m}g_{\a\b}=\o_{\m}g_{\a\b}.
\ee
%In the above
Here $\o_{\m}$ denotes a 1-form field known as gauge vector field,
and $^{(W)}\nabla_{\m}$ is the Weyl covariant derivative whose affine connection (Weyl connection) is
\be
\label{2.2}
^{(W)}\G^{\l}_{\a\b}=\big \{^{\l}_{\a\b}\big\}-\f{1}{2}(\d^{\l}_{\a}\o_{\b}+\d^{\l}_{\b}\o_{\a}-g_{\a\b}g^{\l\s}\o_{\s}),
\ee
with $\big \{^{\l}_{\a\b}\big\}$ the Levi-Civita connection in Riemannian geometry.

Roughly speaking, Weyl's idea to unify gravity with electromagnetism in the framework of Weyl geometry is to interpret the gauge vector field $\o_{\m}$ just as the electromagnetic field.
However, the presence of $\o_{\m}$ causes the length of a vector, i.e. $l=\sqrt {g_{\a\b}l^{\a}l^{\b}}$, to vary point
to point as $\f{dl}{l}=\f{1}{2}\o_{\m}dx^{\m}$.
 Thus, its length will generally take a different value after the vector has been parallel-transported in a closed path: $l=l_0\exp\oint \o_{\m}dx^{\m}$.
 As pointed out by Einstein, this would phenomenally
 lead to a broadening of the atomic spectral lines for electrons immersed in the $\o_{\m}$ field.
Since this so-called "second clock effect" has not been confirmed experimentally,
Weyl geometry was considered not physically viable.

To make a way out of this dilemma, Weyl subsequently proposed a particular class of Weyl geometry later known as Weyl integrable geometry.
This entails that the gauge vector field $\o_{\m}$ is now restricted to be an exact form which can be described by the derivative of a scalar field $\phi$, i.e. $\o_{\m} \equiv \p_{\m}\psi$. Then (\ref{2.1}) and (\ref{2.2}) will take the following form
\be
\label{2.3}
^{(WI)}\nabla_{\m}g_{\a\b}=\p_{\m} \psi\cdot g_{\a\b},
\ee
\be
\la{2.33}
\qquad ^{(WI)}\G^{\l}_{\a\b}=\big \{^{\l}_{\a\b}\big\}-\f{1}{2}(\d^{\l}_{\a}\p_{\b}\psi+\d^{\l}_{\b}\p_{\a}\psi-g_{\a\b}g^{\l\s}\p_{\s}\psi).
\ee
By construction the length of a vector is invariant when it is parallel-transported along a closed path, because the Stokes's theorem ensures that $l=l_0\exp\oint \p_{\m}\psi dx^{\m}=l_0$.

It is obvious that no torsion is present in Weyl and Weyl integrable geometry since the two lower indices of the Weyl connection are symmetric.
Furthermore, an important feature of Weyl geometry is that (\ref{2.1}) and (\ref{2.2}) are invariant under the following simultaneous transformations in $g_{\a\b}$ and $\o_{\m}$, also known as Weyl rescalings or Weyl gauge transformation:
\be
\label{2.4}
\bar g_{\a\b}=\O^2g_{\a\b}, \quad \bar \o_{\m}=\o_{\m}+2\p_{\m}\ln\O.
\ee
Note that besides the conformal transformation, Weyl rescalings also involve the transformation of $\o_{\m}$.
Weyl integrable geometry has the similar characteristic except the Weyl rescalings now take a simpler form as
\be
\label{2.5}
\bar g_{\a\b}=\O^2g_{\a\b}, \qquad \bar \psi=\psi+2\ln\O.
\ee
It can be verified that this connection (\ref{2.33}) is also invariant under Weyl rescalings.

Thus in Weyl (or Weyl integrable) geometry, the metric $g_{\a\b}$ and gauge vector $\o_{\m}$ (or $\p_{\m}\psi$) in fact represent an equivalence class of metrics and gauge vectors. Two given pairs, $(g_{\a\b}, \o_{\m})$ and $(g'_{\a\b}, \o'_{\m})$,  are in the same equivalence class if they are related to each other through Weyl rescalings defined by (\ref{2.4}) or (\ref{2.5}). Thus, for any given Weyl integrable geometry with pair $(g_{\a\b}, \psi)$,
one can always choose $\O=e^{-\f{1}{2}\psi}$ in the Weyl rescalings, then the definition (\ref{2.3}) for Weyl integrable geometry turns into
\be
\label{2.6}
\nabla_{\m}\bar g_{\a\b}=0,
\ee
\be
\la{2.66}
\qquad \G^{\l}_{\a\b}=\big \{^{\bar \l}_{\a\b}\big\}=\f{1}{2}\bar g^{\l\s}(\p_{\a}\bar g_{\b\s}+\p_{\b}\bar g_{\a\s}-\p_{\s}\bar g_{\a\b}),
\ee
which defines a Riemannian geometry. The concern is, since the pairs $(g_{\a\b}, \psi)$ and $(\bar g_{\a\b}= e^{-\psi} g_{\a\b}, \bar \psi=0)$  are in the same equivalence class, they actually define the same geometry but only with different gauges.

\section{Induced geometry from disformal transformation}    \la{sec3}

\subsection{The case of conformal transformation}

Before we settle down to induce geometry from Riemannian geometry by disformal transformation, it's helpful to start with a simpler example for  the clarification of concepts and fixing conventions, thus we want to show how is Weyl integrable geometry induced by conformal transformation here.

It's apparent from Section \ref{sec2} that, from Weyl integrable geometry, one can get a Riemannian geometry by appropriate Weyl rescalings.
However, this process certainly depends on our knowledge about the Weyl integrable geometry. If one has no \emph{a priori} knowledge of the so-called Weyl integrable geometry, can she/he induce it from the mere knowledge about the starting Riemannian geometry?

To answer this question, we start from the metricity condition and corresponding Levi-Civita connection for Riemannian geometry which are
\be
\label{2.7}
\nabla_{\m} g_{\a\b}=0,\qquad \G^{\l}_{\a\b}=\big \{^{\l}_{\a\b}\big\}=\f{1}{2}g^{\l\s}(\p_{\a}g_{\b\s}+\p_{\b}g_{\a\s}-\p_{\s}g_{\a\b}).
\ee

Then, we do conformal transformation to the metric. We stress that this operation should not be treated as new input to the theory, but a change of units \cite{Dicke:1961gz}. For later convenience,
we denote the conformal transformation as
\be
\label{2.8}
\bar g_{\a\b}=A(\phi)g_{\a\b},
\ee
with $A(\phi)$ a positive definite function of scalar field $\phi$.
Now, a simple calculation can be done as follows:
\bea
\label{2.91}
\nabla_{\m}g_{\a\b}&=&0=\nabla_{\m}\big( A^{-1}\bar g_{\a\b}\big) \nn \\
&=&-A^{-2}A_{\m}\bar  g_{\a\b}+A^{-1}\nabla_{\m}\bar g_{\a\b},
\eea
with $A_{\m}=\nabla_{\m}A$. Then, rearranging (\ref{2.91}) appropriately, we arrive at
\be
\label{2.92}
\nabla_{\m}\bar  g_{\a\b}=\p_{\m}(\ln A)\cdot \bar  g_{\a\b}.
\ee
Furthermore, we can also rewrite the Levi-Civita connection for $g_{\a\b}$ in the new field variable $\bar g_{\a\b}$ as
\be
\la{2.93}
\{^{\l}_{\a\b}\}=\{^{\bar \l}_{\a\b}\}-\frac{1}{2}\big [\d^{\l}_{\a}\p_{\b}(\ln A)+\d^{\l}_{\b}\p_{\a}(\ln A)- \bar g_{\a\b} \bar g^{\l\s}\p_{\s}(\ln A)\big ],
\ee
with $\{^{\bar \l}_{\a\b}\}$ the Levi-Civita connection corresponding to $\bar g_{\a\b}$. It's also easy to check that, the new nonmetricity condition and connection are invariant under
simultaneous transformations in $\bar g_{\a\b}$ and $\ln A$:
\be
\la{2.94}
\bar {\bar g}_{\a\b}= B(\varphi)\bar g_{\a\b}, \quad \ln \bar A=\ln A+\ln B.
\ee

Now it's obvious that, if we denote $\psi=\ln A, \quad \O^2=B$ and rewrite $\bar g_{\a\b}$ as $g_{\a\b}$ for convenience, (\ref{2.92})-(\ref{2.94}) are exactly the same as (\ref{2.3}) and (\ref{2.5}) for Weyl integrable geometry. The crucial point is that the new nonmetricity condition (\ref{2.92}) is totally and exclusively induced from the starting metricity condition (\ref{2.7}) for Riemannian geometry without any new input and any new constraint, so is the new connection. It is in this sense that we call Weyl integrable geometry an "induced geometry".

\subsection{Nonmetricity condition and affine connection}

Now we start to induce new geometry from disformal transformation by the method we have demonstrated and investigated in detail. The disformal transformation introduced by Bekenstein \cite{Bekenstein:1992pj} has the following form:
\be
\label{4.1}
\bar g_{\a\b}=A(\phi, X)g_{\a\b}+B(\phi, X)\phi_{\a}\phi_{\b},
\ee
where $X = \f{1}{2} g^{\a\b}\phi_\a \phi_\b$.
Recently it has been shown in \cite{{Bettoni:2013diz},{Zumalacarregui:2013pma}} that Horndeski theories are invariant under the
special disformal transformation
\be
\label{4.222}
\bar g_{\a\b}=A(\phi)g_{\a\b}+B(\phi)\phi_{\a}\phi_{\b}.
\ee
This will be the starting point of our subsequent discussions.

The goal we want to achieve is to derive a new (non)metricity condition from Riemannian geometry. This process must be conducted without any new input or any new constraint to insure the induced geometry is faithfully \textit{induced} from the starting Riemannian geometry by disformal transformation. With these considerations, we treat (\ref{4.2}) as a complicated version of change of units, i.e., it does not introduce new input or new constraint into the theory. Then, we start with two equations which can be derived from (\ref{4.222}) that
\bea
\la{4.2}
g_{\a\b}&=&\f{1}{A}\bar g_{\a\b}-\f{B}{A}\phi_{\a}\phi_{\b}, \\
\la{4.3}
g^{\a\b}&=&A\bar g^{\a\b}+\f{AB}{1-2B\bar X}\bar \phi^{\a}\bar \phi^{\b},
\eea
where $\bar X =\f{1}{2}\phi_{\m}\phi_{\n}\bar g^{\m\n}$ and $\bar \phi^{\m}=\phi_{\n}\bar g^{\n\m}$.

Then, a simple calculation can be done as
\bea
\la{4.4}
\nabla_{\m}g_{\a\b}&=&0 =\nabla_{\m}(\f{1}{A}\bar g_{\a\b}-\f{B}{A}\phi_{\a}\phi_{\b})  \nn \\
&=&-A^{-2}A_{\m}\bar g_{\a\b}+\f{1}{A}\nabla_{\m}\bar g_{\a\b}-\nabla_{\m}(\f{B}{A}\phi_{\a}\phi_{\b}).
\eea
Rearranging the results as
\be
\la{4.5}
\nabla_{\m}\bar g_{\a\b}=\f{A_{\m}}{A}\bar g_{\a\b}+A\nabla_{\m}(\f{B}{A}\phi_{\a}\phi_{\b}).
\ee
It is obvious that, if we restrict to disformal transformation with $B=0$, (\ref{4.4}) turns right back to (\ref{2.92}) which is for induced geometry from conformal transformation.
(\ref{4.4}) looks like a nonmetricity condition which is of our expectation. Nevertheless, in order to get the induced geometry, we need to rewrite both sides of this equation with new field variables $\bar g_{\a\b}$ and $\phi_{\a}$, this in fact comes down to rewrite the original Levi-Civita connection in new variables. This can be done by inserting (\ref{4.3}) into the connection, then after some calculation we can arrive at
\bea
\la{4.6}
\big \{^{\l}_{\a\b}\big\}&=&\big \{^{\bar \l}_{\a\b}\big\}-\f{A'}{2A}(\d^{\l}_{\b}\phi_{\a}+\d^{\l}_{\a}\phi_{\b})+\f{A'}{2A(1-2B\bar X)}\bar \phi^{\l}\bar g_{\a\b} \nn \\
&&-\f{A'B+AB'}{2A(1-2B\bar X)}\bar \phi^{\l}\phi_{\a}\phi_{\b}-\f{B}{2(1-2B\bar X)}\bar \phi^{\l}(\bar \phi_{\a\b}+\bar \phi_{\b\a}),
\eea
with the prime denoting the derivative with respect to $\phi$ and $\bar \phi_{\a\b}=\p_{\b}\phi_{\a}-\big \{^{\bar \l}_{\a\b}\big \}\phi_{\l}$.

The tricky part of this story is that if one starts with \cite{Bettoni:2013diz}
\bea
\la{4.7}
\big \{^{\l}_{\a\b}\big\}&=&\big \{^{\bar \l}_{\a\b}\big\}-\f{A'}{2A}(\d^{\l}_{\b}\phi_{\a}+\d^{\l}_{\a}\phi_{\b})-\f{1}{2}\f{\phi^{\l}}{A^2(1+2XB/A)}\big (-AA'g_{\a\b}+(AB'-2A'B)\phi_{\a}\phi_{\b}\big ) \nn \\
&&-\f{B}{A(1+2XB/A)}\phi^{\l}\phi_{\a\b},
\eea
then use $g^{\a\b}=A\bar g^{\a\b}+\f{AB}{1-2B\bar X}\bar \phi^{\a}\bar \phi^{\b}$ and $X=\f{A\bar X}{1-2B\bar X}$ to rewrite the original metric in (\ref{4.7}) into new metric, it's in fact still difficult to get a formula which is described completely by $\bar g_{\a\b}$ and $\phi$, for there is always a Levi-Civita connection for the original metric $g_{\a\b}$ in term $\phi_{\a\b}$ which causes the rewriting of field variables into an infinite iteration. The more economical, or even the only way, to arrive the formula (\ref{4.6}) is to proceed the calculation done as here. This is totally different from what has happened for inducing geometry from Riemannian geometry by conformal transformation, which also implies that disformal transformation is a nontrivial generalization of conformal transformation.

Now inserting
 (\ref{4.6}) into the RHS of (\ref {4.5}) and rewriting $\bar g_{\a\b}$ as $g_{\a\b}$ for convenience, we arrive at the new induced nonmetricity condition
\bea
\la{4.8}
\nabla_{\m} g_{\a\b}&=&\f{A_{\m}}{A} g_{\a\b}-\f{A'B X}{A(1-2B X)}( g_{\a\m}\phi_{\b}+ g_{\b\m}\phi_{\a})+\f{A'B+AB'}{A(1-2B X)}\phi_{\a}\phi_{\b}\phi_{\m} \nn \\
&&+\f{B}{1-2B X}(\phi_{\b} \phi_{\a\m}+\phi_{\a} \phi_{\b\m}),
\eea
with the new induced affine connection defined as
\bea
\la{4.9}
\G^{\l}_{\a\b}&=&\big \{^{\l}_{\a\b}\big\}-\f{A'}{2A}(\d^{\l}_{\b}\phi_{\a}+\d^{\l}_{\a}\phi_{\b})+\f{A'}{2A(1-2B X)} \phi^{\l}g_{\a\b} \nn \\
&&-\f{A'B+AB'}{2A(1-2B X)}\phi^{\l}\phi_{\a}\phi_{\b}-\f{B}{2(1-2B X)}\phi^{\l}(\phi_{\a\b}+\phi_{\b\a}).
\eea
One should notice that $\phi_{\a\b}$ is understood as  $\phi_{\a\b}
=\p_{\b}\phi_{\a}-\big \{^{\l}_{\a\b}\big\}\phi_{\l}$.

These two formulas, (\ref{4.8}) and (\ref{4.9}) are just what we look for to define a new geometry. They are both induced from Riemannian geometry by disformal transformation. The disformal transformation of the metric is just a generalized change of units, thus this operation on the field variable of the theory does not introduce any new input or new constraint. Exactly because of this reason  we call our finding an "induced geometry" from disformal transformation.

\subsection{Gauge transformation}

For a special disformal transformation of the original metric $\bar g_{\a\b}=Ag_{\a\b}+B\phi_{\a}\phi_{\b}$,
we have
\be \la{4.10}
\nabla_{\m}\bar g_{\a\b}=\f{A_{\m}}{A}\bar g_{\a\b}+\nabla_{\m}(B\phi_{\a}\phi_{\b})-\f{A_{\m}}{A}B\phi_{\a}\phi_{\b}.
\ee
Now we recall that the gauge transformation for metric in Weyl integrable geometry, which is an induced geometry from conformal transformation, is a conformal transformation of the original metric. Thus, it is reasonable to infer that the gauge transformation for metric used to define geometry induced from disformal transformation, is also a disformal transformation. In fact, if we implement the second disformal transformation to the metric $\bar g_{\a\b}$ as
\bea
\label{4.11}
\bar {\bar g}_{\a\b}&=&C\bar g_{\a\b}+D\psi_{\a}\psi_{\b} \nn \\
&=&ACg_{\a\b}+BC\phi_{\a}\phi_{\b}+D\psi_{\a}\psi_{\b}.
\eea
we arrive at
\be
\label{4.12}
\nabla_{\m}\bar {\bar g}_{\a\b}=\f{(AC)_{\m}}{AC}\bar {\bar g}_{\a\b}+\nabla_{\m}(BC\phi_{\a}\phi_{\b}+D\psi_{\a}\psi_{\b})-\f{(AC)_{\m}}{AC}(BC\phi_{\a}\phi_{\b}+D\psi_{\a}\psi_{\b}).
\ee
Through the comparison of (\ref{4.10}) and (\ref{4.12}), it can be seen that the nonmetricity condition (\ref{4.12}) is invariant under the following simultaneous transformations:
\be
\la{4.133}
\begin{cases}
\bar g_{\a\b}\longrightarrow \bar {\bar g}_{\a\b}= ACg_{\a\b}+BC\phi_{\a}\phi_{\b}+D\psi_{\a}\psi_{\b}, \\
A \longrightarrow AC, \\
B\phi_{\a}\phi_{\b} \longrightarrow BC\phi_{\a}\phi_{\b}+D\psi_{\a}\psi_{\b}.
\end{cases}
\ee
Then, a further deduction can be made is that, since it is totally induced from (\ref{4.5}) which is equivalent to (\ref{4.10}), they must be invariant under the same gauge transformation. Thus, it seems that (\ref{4.133}) is indeed the expected gauge transformation for our induced geometry.

Nevertheless, it should be stressed here that (\ref{4.133}) includes only a subset of the total allowed gauge transformations. To get the whole set of gauge transformation for the induced geometry, we first rewrite (\ref{4.11}) as
\be
\la{4.22}
g_{\a\b}=\f{1}{AC}\bar {\bar g}_{\a\b}-\f{B}{A}\phi_{\a}\phi_{\b}-\f{D}{AC}\psi_{\a}\psi_{\b}.
\ee
Comparing this with (\ref{4.2}) and (\ref{4.3}), it seems that one can expect that the inverse metric $g^{\a\b}$ can take the form as
\be
\la{4.23}
g^{\a\b}=E\bar {\bar g}^{\a\b}+F\bar {\bar \phi}^{\a}\bar {\bar \phi}^{\b}+G\bar {\bar \psi}^{\a}\bar {\bar \psi}^{\b}.
\ee
However, a detailed investigation shows that, generally, there is no solutions to the constraint equations set by $g_{\a\b}g^{\a\r}=\d^{\r}_{\b}$, which tells that the inverse metric in form of (\ref{4.23}) does not exist. This will inevitably forbid a successful rewriting of the Levi-Civita connection in (\ref{4.12}) with disformal-transformed metric $\bar {\bar g}_{\a\b}$, thus makes it impossible to define the correct connection for the induced geometry.

The way out of this dilemma is to notice the fact that we do not take all reasonable terms into consideration in (\ref{4.23}). The term which should not be ignored is $H(\bar {\bar \psi}^{\a}\bar {\bar \phi}^{\b}+\bar {\bar \phi}^{\a}\bar {\bar \psi}^{\b})$. If we supplement this term into (\ref{4.23}) and solve constraint equations set by $g_{\a\b}g^{\a\r}=\d^{\r}_{\b}$, we can find that there are indeed consistent solutions about $E,F,G,H$. More importantly, this fact in turn inspires us that the whole gauge transformation of the metric should be defined as
\bea
\la{4.24}
\bar {\bar g}_{\a\b}&=&C\bar g_{\a\b}+D\psi_{\a}\psi_{\b}+I(\phi_{\a}\psi_{\b}+\psi_{\a}\phi_{\b}) \nn \\
&=&ACg_{\a\b}+BC\phi_{\a}\phi_{\b}+D\psi_{\a}\psi_{\b}+I(\phi_{\a}\psi_{\b}+\psi_{\a}\phi_{\b}).
\eea
Through some tedious but straight calculation, one can show that, the original metric $g_{\a\b}$ and its inverse can be rewrite with new metric as
\bea
\la{4.25}
g_{\a\b}&=&\f{1}{AC}\bar {\bar g}_{\a\b}-\f{B}{A}\phi_{\a}\phi_{\b}-\f{D}{AC}\psi_{\a}\psi_{\b}-\f{I}{AC}(\phi_{\a}\psi_{\b}+\psi_{\a}\phi_{\b}), \\
\la{4.26}
g^{\a\b}&=&AC\bar {\bar g}^{\a\b}+\f{AC}{ac-bd}\big [(BCc+Ib)\bar {\bar \phi}^{\a}\bar {\bar \phi}^{\b}+(Da-Ib)\bar {\bar \psi}^{\a}\bar {\bar \psi}^{\b} \nn \\
&&\qquad \qquad \qquad \qquad \qquad \qquad \qquad+(BCd+Ia)(\bar {\bar \phi}^{\a}\bar {\bar \psi}^{\b}+\bar {\bar \psi}^{\a}\bar {\bar \phi}^{\b})\big ],
\eea
with $a,b,c,d$ defined as
\be
\label{4.27}
\begin{cases}
a=1-2BC\bar {\bar X}-2I\bar {\bar Z}, \\
b=2BC\bar {\bar Z}+2I\bar {\bar Y}, \\
c=1-2D\bar {\bar Y}-2I\bar {\bar Z}, \\
d=2D\bar {\bar Z}+2I\bar {\bar X},
\end{cases}
\ee
where $\bar {\bar X}=\f{1}{2}\phi_{\m}\phi_{\n}\bar {\bar g}^{\m\n}, \bar {\bar Y}=\f{1}{2}\psi_{\m}\psi_{\n}\bar {\bar g}^{\m\n}$ and $\bar {\bar Z}=\f{1}{2}\phi_{\m}\psi_{\n}\bar {\bar g}^{\m\n}$. With these results in hand, we can rewrite the original Levi-Civita connection with new variables and interpret it as the induced connection for the induced geometry. Thus, transformation defined by (\ref{4.24}) is indeed the gauge transformation for our induced geometry. The fascinating feature of this transformation is that it is not a trivial expected disformal transformation but with additional crossover term, $I(\phi_{\a}\psi_{\b}+\psi_{\a}\phi_{\b})$.

For clarity, we rewrite $\bar {\bar g}_{\a\b}$ as $\bar g_{\a\b}$ in (\ref{4.24}) to keep correspondence with the notation used in (\ref{4.8}) and (\ref{4.9}), then, the gauge transformation of the induced geometry can be extracted from (\ref{4.24}) as follows:
\be
\label{4.28}
\begin{cases}
g_{\a\b}\longrightarrow \bar g_{\a\b}= ACg_{\a\b}+BC\phi_{\a}\phi_{\b}+D\psi_{\a}\psi_{\b}+I(\phi_{\a}\psi_{\b}+\psi_{\a}\phi_{\b}),\\
A \longrightarrow AC, \\
B\phi_{\a}\phi_{\b} \longrightarrow BC\phi_{\a}\phi_{\b}+D\psi_{\a}\psi_{\b}+I(\phi_{\a}\psi_{\b}+\psi_{\a}\phi_{\b}).
\end{cases}
\ee

\subsection{Generalized Weyl geometry}

The induced geometry defined by (\ref{4.8}) and (\ref{4.9}) can be naturally considered as a generalization of Weyl integrable geometry. Recalling the fact that Weyl geometry can be easily got by replacing $\p_{\m}\phi$ in (\ref{2.3}) and (\ref{2.4}), which are used to define Weyl integrable geometry, with 1-form field $\o_{\m}$, we can expect that implementing similar replacement in (\ref{4.8}) and (\ref{4.9}) will leads to new geometry which is a natural generalization of Weyl geometry. In order to do this, we first rewrite (\ref{4.8}) as
\bea
\la{4.14}
\nabla_{\m} g_{\a\b}&=&\p_{\m}(\ln A)( g_{\a\b}+B\phi_{\a}\phi_{\b})- \f{B X\p_{\b}(\ln A)}{1-2B X}( g_{\a\m}-B\phi_{\a}\phi_{\m}) \nn \\
&&-\f{B X\p_{\a}(\ln A)}{1-2B X}( g_{\b\m}-B\phi_{\b}\phi_{\m})+\f{1}{1-2B X}(B\phi_{\a}\phi_{\b})_{;\m}.
\eea
We also rearrange (\ref{4.9}) as
\bea
\la{4.15}
\G^{\l}_{\a\b}&=&\big \{^{\l}_{\a\b}\big\}-\f{1}{2}\big (\d^{\l}_{\b}\p_{\a}(\ln A)+\d^{\l}_{\a}\p_{\b}(\ln A)\big )+\f{\p^{\l}(\ln A)}{2(1-2B X)}(g_{\a\b}-B\phi_{\a}\phi_{\b}) \nn \\
&&-\f{\phi^{\l}}{4(1-2B X)}(B_{\b}\phi_{\a}+B_{\a}\phi_{\b})-\f{B\phi^{\l}}{2(1-2B X)}(\phi_{\a\b}+\phi_{\b\a}).
\eea

Now, we would like to replace $\p_{\m}(\ln A)$ and $\phi_{\a}$ with 1-form field $\o_{\m}$ and $\chi_{\a}$ respectively, and treat $X$ as the norm of the 1-form field $\chi_{\a}$. A careful treatment of term $\f{B_{\b}}{2(1-2B X)}\phi^{\l}\phi_{\a}$ is needed. In (\ref{4.8}) and (\ref{4.9}), $B$ is defined as a scalar functional of the scalar field $\phi$ and $B_{\b}=\f{dB}{d\phi}\f{\p \phi}{\p x^{\b}}$. However, after the aforementioned replacement is made, the restriction on $B$ and $B_{\b}$ does not make sense anymore, thus the meaning of $B_{\b}$ needs to be specified individually. In order to be as general as possible, we would like to require that $B$ is not a scalar functional of other field anymore but a scalar field only, and $B_{\b}$ is just its gradient.

Then, the generalized geometry corresponding to Weyl geometry is defined by nonmetricity condition
\bea
\la{4.16}
\nabla_{\m} g_{\a\b}&=&\o_{\m}( g_{\a\b}+B\chi_{\a}\chi_{\b})- \f{B X\o_{\b}}{1-2B X}( g_{\a\m}-B\chi_{\a}\chi_{\m}) \nn \\
&&-\f{B X\o_{\a}}{1-2B X}( g_{\b\m}-B\chi_{\b}\chi_{\m})+\f{1}{1-2B X}(B\chi_{\a}\chi_{\b})_{;\m},
\eea
with its connection defined as
\bea
\la{4.17}
\G^{\l}_{\a\b}&=&\big \{^{\l}_{\a\b}\big\}-\f{1}{2}\big (\d^{\l}_{\b}\o_{\a}+\d^{\l}_{\a}\o_{\b}\big )+\f{\o^{\l}}{2(1-2B X)}(g_{\a\b}-B\chi_{\a}\chi_{\b}) \nn \\
&&-\f{\chi^{\l}}{4(1-2B X)}(B_{\b}\chi_{\a}+B_{\a}\chi_{\b})-\f{B\chi^{\l}}{2(1-2B X)}(\chi_{\a;\b}+\chi_{\b;\a}).
\eea

\section{Conclusion and discussion}

For a conclusion, we start from Riemannian geometry and then implement disformal transformation on the metric. By treating this operation as a gauge transformation therefore no new input or constraint is introduced, we induce new nonmetricity condition and connection to define a new geometry, see (\ref{4.8}) and (\ref{4.9}). It is in this sense that we say the new geometry is \textit{induced}.

The important feature of the induced geometry is that it preserves the affine structure of the original Riemannian geometry. No matter what geometry (the original or the induced) is under consideration,  the covariant derivative is in fact the same one, but described with different variables. Concretely, in Riemannian geometry, the Levi-Civita connection is just itself and described by metric $g_{\a\b}$ only; in the induced Weyl integrable geometry, the original Levi-Civita connection is described by the new metric $\bar g_{\a\b}$ and $A$.

Furthermore, noticing the fact that Weyl geometry can be obtained simply by replacing the gradient of the scalar field in Weyl integrable geometry with a 1-form field, we implement similar replacement on (\ref{4.8}) and (\ref{4.9}) to generalize the new induced geometry further. This operation naturally leads to a new geometry which corresponds to a generalization of Weyl geometry, see (\ref{4.16}) and (\ref{4.17}).

A natural generalization of this work would be to consider the usual disformal transformation (\ref{4.1})
or even the extended disformal transformation (with a rank-two symmetric tensor $E_{(\a\b)}$) \cite{Zumalacarregui:2013pma}.
One may also establish a disformal equivalence principle along the line of \cite{Quiros:2011wb}.
Note that in his original work \cite{Bekenstein:1992pj}, Bekenstein invoked Finsler geometry to motivate the introduction of disformal transformation.
It is natural to anticipate that our discussions here may have connections with the mathematical aspects of the Finsler geometry.

\section*{Acknowledgments}
We would like to thank Miao Li and Zhenhui Zhang for helpful discussions.

\end{document}